\documentclass[sigconf]{acmart}

\AtBeginDocument{%
  \providecommand\BibTeX{{%
    \normalfont B\kern-0.5em{\scshape i\kern-0.25em b}\kern-0.8em\TeX}}}

\newcommand{\ie}{\emph{i.e., }}
\newcommand{\eg}{\emph{e.g., }}

\newcommand{\etc}{\emph{etc.}}
\newcommand{\wrt}{\emph{w.r.t. }}
\newcommand{\cf}{\emph{cf. }}
\newcommand{\aka}{\emph{a.k.a. }}
\newcommand{\zy}[1]{#1}

\usepackage{enumitem}
\usepackage{verbatim}
\usepackage[ruled]{algorithm2e}
\usepackage{multirow}
\usepackage{subfigure} 
\usepackage{ragged2e}
\usepackage{caption}
\usepackage{graphicx}
\usepackage[normalem]{ulem}
\useunder{\uline}{\ul}{}
\usepackage{amsmath}
\usepackage{setspace}
\usepackage{bm}

\copyrightyear{2023} 
\acmYear{2023} 
\setcopyright{acmlicensed}\acmConference[SIGIR '23]{Proceedings of the 46th
International ACM SIGIR Conference on Research and Development in
Information Retrieval}{July 23--27, 2023}{Taipei, Taiwan}
\acmBooktitle{Proceedings of the 46th International ACM SIGIR Conference on
Research and Development in Information Retrieval (SIGIR '23), July 23--27,
2023, Taipei, Taiwan}
\acmPrice{15.00}
\acmDOI{10.1145/3539618.3591755}
\acmISBN{978-1-4503-9408-6/23/07}

\begin{document}

\title{Reformulating CTR Prediction: Learning Invariant Feature Interactions for Recommendation}

\author{Yang Zhang}
\authornote{Equal contribution.}
\affiliation{%
	\institution{University of Science and Technology of China}
	\city{Hefei}
	\country{China}}
\email{zy2015@mail.ustc.edu.com}

\author{Tianhao Shi}
\authornotemark[1]
\affiliation{%
	\institution{University of Science and Technology of China}
	\city{Hefei}
	\country{China}}
\email{sth@mail.ustc.edu.com}

\author{Fuli Feng}
\authornote{Corresponding authors.}
\affiliation{%
	\institution{University of Science and Technology of China}
	\city{Hefei}
	\country{China}}
\email{fulifeng93@gmail.com}

\author{Wenjie Wang}
\affiliation{%
	\institution{National University of Singapore}
	\country{Singapore}}
\email{wenjiewang96@gmail.com}

\author{Dingxian Wang}
\affiliation{%
	\institution{Etsy Inc.}
	\country{United States}
}
\email{dingxianwang@etsy.com}

\author{Xiangnan He}
\authornotemark[2]
\affiliation{%
	\institution{University of Science and Technology of China}
	\city{Hefei}
	\country{China}}
\email{xiangnanhe@gmail.com}

\author{Yongdong Zhang}
\affiliation{%
	\institution{University of Science and Technology of China}
	\city{Hefei}
	\country{China}}
\email{zhyd73@ustc.edu.cn}
 






\renewcommand{\shortauthors}{Yang Zhang, et al.}

\begin{abstract}

Click-Through Rate (CTR) prediction plays a core role in recommender systems, serving as the final-stage filter to rank items for a user. The key to addressing the CTR task is learning feature interactions that are useful for prediction, which is typically achieved by fitting historical click data with the Empirical Risk Minimization (ERM) paradigm. Representative methods include Factorization Machines and Deep Interest Network, which have achieved wide success in industrial applications. However, such a manner inevitably learns \textit{unstable feature interactions}, \ie the ones that exhibit strong correlations in historical data but generalize poorly for future serving. 

In this work, we reformulate the CTR task --- instead of pursuing ERM on historical data, we split the historical data chronologically into several periods (\aka environments), aiming to learn feature interactions that are \textbf{stable} across periods. Such feature interactions are supposed to generalize better to predict future behavior data. Nevertheless, a technical challenge is that existing invariant learning solutions like Invariant Risk Minimization are not applicable, since the click data entangles both environment-invariant and environment-specific correlations. To address this dilemma, we propose \textit{Disentangled Invariant Learning} (DIL) which disentangles feature embeddings to capture the two types of correlations separately. To improve the modeling efficiency, we further design LightDIL which performs the disentanglement at the higher level of the feature field. Extensive experiments demonstrate the effectiveness of DIL in learning stable feature interactions for CTR. We release the code at https://github.com/zyang1580/DIL.

\end{abstract}




\maketitle
\section{Introduction}

Click-Through Rate (CTR) prediction is important to support the ranking stage of recommender systems~\cite{ctrsurvey}.
Its key lies in learning the feature interactions that are useful for predicting user clicks~\cite{ctrsurvey, DeepFM,nas-ctr}. Existing methods~\cite{ctrsurvey, FM, NFM,DeepFM,DIN,can,min2022neighbour} typically achieve the target by fitting historical data collectively with the Empirical Risk Minimization (ERM) paradigm ---  minimizing the average empirical loss over historical click data~\cite{ctrsurvey}. The ERM paradigm has become the standard paradigm for training CTR models, leading to the classical solutions FM~\cite{FM}, NFM~\cite{NFM}, and DIN~\cite{DIN} that have been intensively used in industrial applications~\cite{ctrsurvey}. 

\begin{figure}[t]
  \centering
  \includegraphics[width=\linewidth]{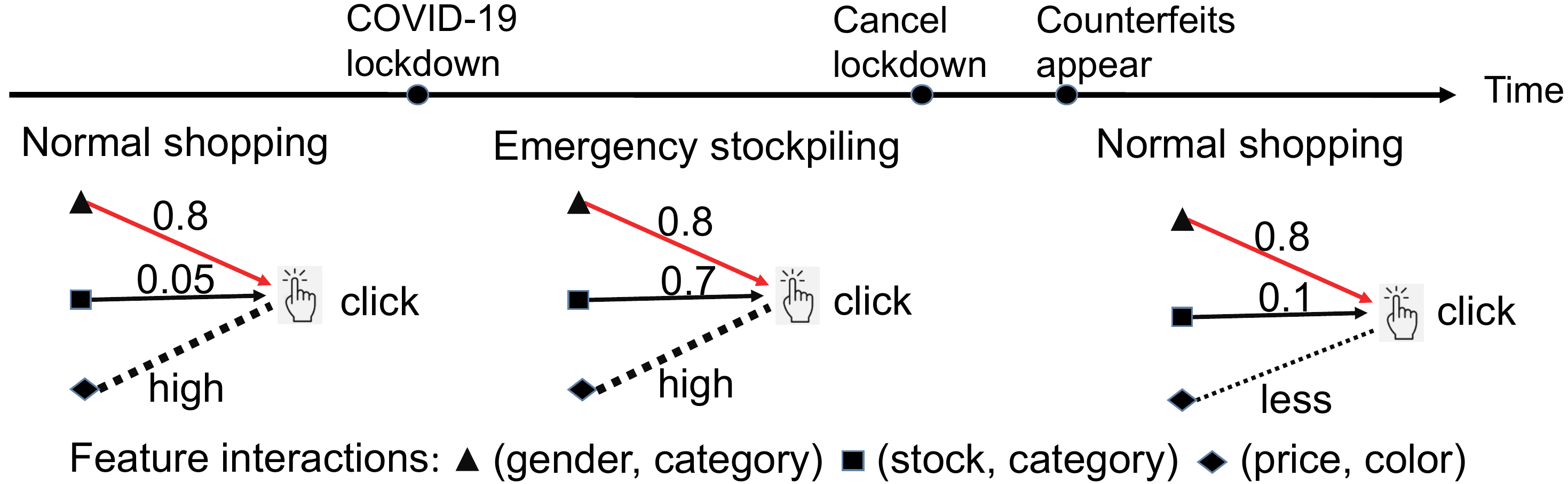}
    \vspace{-17pt}
  \caption{
  Examples of stable feature interaction (gender, category) and unstable feature interaction (stock, category) and (price, color).
  The arrow represents that the feature interaction affects the happening of click, and the weights reflect the influence strength. The dotted line represents the correlations between feature interactions and clicks, varying along the timeline.
  }
  \vspace{-15pt}
  \label{fig:example}
\end{figure}

We argue that the CTR model built with the ERM paradigm would inevitably learn \textit{unstable feature interactions}, leading to poor generalization for future serving. Some feature interactions could exhibit strong correlations in historical data but are useless or even harmful for future serving. 
Figure 1 illustrates two examples of unstable feature interactions: 1) the feature interaction (stock, category) highly affects user clicks during the COVID-19 lockdown, however, its effect diminishes after the lockdown; and 2) the feature interaction (price, color) exhibits high correlation with clicks due to some confounding effect\footnote{E.g., the high quality of the product could synchronously lead to high item price, attractive item color, and strong click probability.}, which is however unstable and becomes weak after the occurrence of counterfeits. 
These unstable feature interactions will mislead the CTR model learned by ERM, \eg 1) over-recommending emergent supplies after the lockdown; and 2) over-recommending the counterfeits with high prices and attractive colors.
It is thus essential to distinguish unstable feature interactions in CTR prediction.

In this work, we propose to reformulate the CTR task instead of performing ERM on historical data. As unstable feature interactions could cause poor generalization, we set the target as learning the feature interactions that are stably useful for predicting user clicks.
Towards the goal, we split the historical data chronologically into several periods, forming a set of environments, and learn feature interactions that are stable across these environments.
These environments could be heterogeneous due to the temporal influence, \eg some environments have more historical data during the COVID-19 lockdown. 
The feature interactions that are stably useful over these heterogeneous environments are more likely to remain useful in the future~\cite{IRM,HRM}, \eg the interaction (gender, category) in Figure 1. As such, the CTR models accounting for the stability of feature interactions will generalize better for future prediction.

Among existing techniques, Invariant Learning~\cite{IRM} is well-known for capturing environment-invariant correlations across heterogeneous environments, which seems to be a promising choice for learning stable feature interactions in CTR prediction. However, it is not directly applicable,
because it relies on the sufficiency prediction condition --- the prediction target can be sufficiently predicted with only environment-invariant correlations --- which is not satisfied in CTR prediction. 
Although unstable, some feature interactions truly affect the clicks in a specific environment, \eg during the COVID-19 lockdown, users click items by considering the (stock, category) factor. 
In other words, the clicks entangle both environment-invariant and environment-specific correlations, meaning that the clicks cannot be sufficiently predicted with only environment-invariant correlations.

To overcome this challenge, we propose \textit{Disentangled Invariant Learning} (DIL), which disentangles the environment-invariant and environment-specific correlations from clicks, and identifies stable feature interactions by avoiding the influence of environment-specific correlations. 
Specifically, DIL first equips a CTR model with environment-invariant feature embedding and a set of environment-specific feature embeddings. 
Then DIL optimizes the model with a new disentangled invariant learning objective, which contains: 1) environment-specific objectives to make environment-specific feature embeddings capture environment-specific correlations, and 2) a modified invariant learning objective with the sufficiency prediction condition satisfied to let environment-invariant feature embeddings capture environment-invariant correlations.
Considering that embedding disentanglement significantly increases the model size and the cost of training, we further design LightDIL to perform the disentanglement at the higher level of the feature field to improve the modeling efficiency.

The main contributions are summarized as follows:
\begin{itemize}[leftmargin=*]
    \item New Problem: It is the first time that the CTR task is reformulated by learning stable feature interactions across periods, so as to achieve better generalization for future serving.
    
    \item New Technique: We propose to integrate the idea of representation disentanglement into invariant learning, making invariant learning feasible in difficult situations where the sufficiency prediction condition is not satisfied.
    
    
    
    \item  Experiments: We conduct extensive experiments on both semi-synthetic and real-world datasets, verifying the effectiveness of our proposal.
\end{itemize}
\vspace{-5pt}
\section{Preliminaries}


This work studies the CTR task on chronologically collected data.
Assuming the data is collected from $T$ periods, which could be days, weeks, or self-defined time spans. 
Let $\mathcal{D}=\{\mathcal{D}_{1},\dots,\mathcal{D}_{T}\}$ denote the data, where each $\mathcal{D}_t$ represents the data collected at the $t$-th period.  
We denote each sample in $\mathcal{D}$ as $(\bm{x}, y)$, where $y \in \{0, 1\}$ represents the click and $\bm{x} = [x_1, \cdots, x_N]^T \in \{0, 1\}^N$ represents the features describing the user-item pair.
$\bm{x}$ is a highly sparse multi-hot vector with a dimensionality of $N$. 
We assume the feature vector contains $M$ fields (\eg age and category) where each field is encoded by one-hot encoding or multi-hot encoding. The target is to learn a CTR model from $\mathcal{D}$ to serve for future periods. In this work, we propose to perform stable feature interaction learning to enhance the generalization of the classic FM model. Next, we introduce the backbone FM model and the basics of invariant learning. 

\vspace{-5pt}
\subsection{Factorization Machines}
FM and its variants have achieved wide success in CTR, owing to the effectiveness of the inner product operation in explicitly modeling feature interactions~\cite{FM,DeepFM,NFM}.
Without losing generality, we take FM as the base model to study stable feature interaction modeling.
Given an input sample with features $\bm{x}=[x_1,\dots,x_N]^T$, FM generates the prediction by modeling all second-order feature interactions via inner product on pairwise feature embeddings: 
\begin{equation}
\vspace{-1pt}
\label{eq:fm}\small
    \hat{y} = \sum_{i=1}^{N}\sum_{j>i}^{N} \langle \bm{v}_{i}, \bm{v}_{j} \rangle \cdot x_{i} x_{j}, 
\vspace{-1pt}
\end{equation}
where $\hat{y}$ denotes the prediction of CTR;
$\bm{v}_{i}$/$\bm{v}_{j}$ denotes the embedding for $i$-th/$j$-th feature, and $\langle \bm{v}_{i}, \bm{v}_{j} \rangle$ refers to the inner product between the two feature embeddings to reconstruct the interaction effect. 
The prediction accounts for the interaction between features $x_i$ and $x_j$ if both features occur, \ie $x_{i} x_{j} = 1$.

\noindent\textbf{$\bullet$ Field-level.} Since each feature belongs to one field, we can also model the interaction at the field level. Formally,
\begin{equation} 
\vspace{-1pt}
\label{eq:fm-field}\small
    \hat{y} = \sum_{i=1}^{M}\sum_{j>i}^{M} \langle \bm{u}_{i}, \bm{u}_{j} \rangle,
\vspace{-1pt}
\end{equation}
where $\bm{u}_i$ and $\bm{u}_{j}$ denote the representation of the $i$-th and $j$-th fields. 
When all fields are one-hot encoding, field-level interaction modeling is equivalent to Equation~\eqref{eq:fm}.
Let $\mathcal{S}_i$ denote the feature indices belonging to the $i$-th field, we can generate the field representation $\bm{u}_{i}$ based on the feature embeddings $\{\bm{v}_k\}_{k\in \mathcal{S}_{i}} $ as $\bm{u}_{i} = \frac{1}{\sum_{k\in \mathcal{S}_{i}}{x_k}} \sum_{k\in \mathcal{S}_{i}} x_{k} \cdot \bm{v}_{k}$.




\subsection{Invariant Learning}
Invariant learning~\cite{ood-survey} is widely used to learn predictors for Out-of-Distribution (OOD) generalization, given the data collected from different environments.
Existing invariant learning methods typically have the following assumption~\cite{EERM, HRM, IRM} (\zy{here, we reuse $t$ to denote an environment since a period represents an environment}):
\newtheorem{assumption}{\textbf{ Assumption}}
\begin{assumption}\label{assump}
There exist random variables 
$\bm{r}$ that satisfy:
\begin{itemize}[leftmargin=*]
    \item \textit{Invariance condition}: for any two environments
    $t$ and $t^{\prime}$, $P(y|\bm{r},t)=P(y|\bm{r},t^{\prime})$ holds, where $y$ denotes the prediction target, \eg the click.
    \item \textit{Sufficiency condition}: $y=f(\bm{r})+\epsilon$, where $f$ represents a function and $\epsilon$ is an independent noise.
\end{itemize}
\end{assumption}

This assumption means that capturing the correlations between the predictor variables $\bm{r}$ and the prediction target $y$ can result in a desired model with good generalization to \textit{unknown} distributions~\cite{krueger2021out}, because these correlations are invariant and sufficient for predicting the target across environments. As such, they are also denoted as \textit{environment-invariant correlations}. Under this assumption, to make a model capture such correlations, one representative method is the variance-based regularizer~\cite{HRM,EERM}, which can be formulated as follows:
\begin{equation}\small \label{eq:IL}
\vspace{-2pt}
     min_{\phi}  \quad \frac{1}{T}\sum_{t=1}^{T} R^{t} + \lambda V_{R},
     \vspace{-1pt}
\end{equation} 
where $\phi$ 
denotes the model parameters, $R^{t}$ represents the loss for the environment $t$, $T$ represents the number of environments, $\lambda$ is a hyper-parameter to control the strength of $V_{R}$, and $V_{R}$ denotes the variance of the loss~\cite{EERM} or the variance of the gradient of the loss over environments~\cite{HRM}. In this work, we adopt the variance of the loss with the consideration of computation efficiency, which is computed as follows:
\vspace{-5pt}
\begin{equation} \label{eq:var}\small
    V_{R} = \frac{1}{T-1} \sum_{t=1}^{T} \left(
        R^{t} - \frac{1}{T}\sum_{t^{\prime}=1}^{T} R^{t^{\prime}}
    \right)^2.
\end{equation}
The key consideration of the method is that environment-invariant correlations should enable better performance across environments.
\section{Methodology}
Given the possible existence of unstable feature interaction, we first rethink the CTR prediction task to shed more light on the new problem setting. We then elaborate on the proposed disentangled invariant learning solution. 

\begin{figure}[t]
  \centering
  \includegraphics[width=\linewidth]{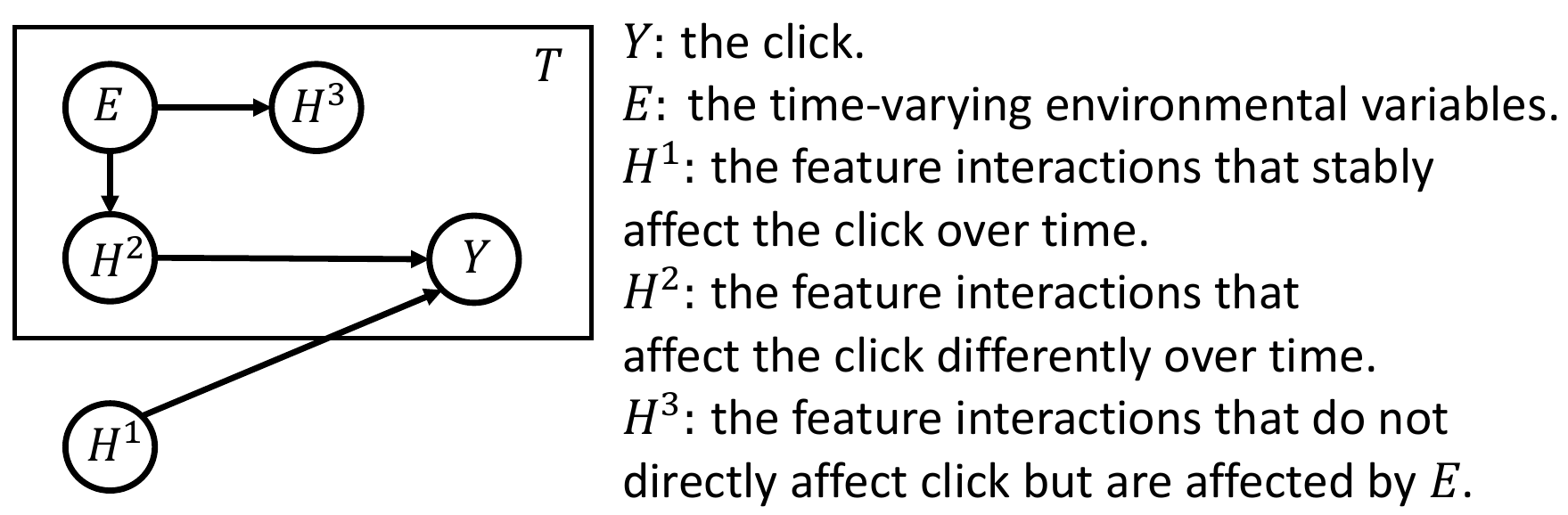}
  \vspace{-16pt}
  \caption{\zy{Graphical model of the click generation process. The box represents $T$ replicates of the part of the model in the box, indicating that the part varies over $T$ periods.}}
  \label{fig:causal-graph}
  \vspace{-16pt}
\end{figure}


\vspace{-3pt}
\subsection{Rethinking the CTR Task} \label{sec:rethink}

%
%
%

\zy{Figure~\ref{fig:causal-graph} illustrates the generative process of click data with a graphical model. The box in the figure represents $T$ replicates of the content in the box, indicating that the content could vary over $T$ periods, \eg $H^2 \rightarrow Y$ 
could vary over $T$ periods. 
We divide the feature interactions into three major types $H^{1}$, $H^2$, $H^3$:}
\begin{itemize} [leftmargin=*]
\item \zy{$H_1$ denotes the feature interactions that stably affect the click over time, as shown by unvarying $H^1 \rightarrow Y$ outside the box.} 


\item \zy{$H_2$ denotes the feature interactions such as (stock, category) that affect the click differently over time due to the influence of time-varying environmental variable $E$ (\eg the COVID-19 lockdown), as shown by time-varying $E \rightarrow H^{2} \rightarrow Y$ in the box.}


\item \zy{$H^3$ denotes the feature interactions that do not directly affect the click $Y$ but are affected by $E$ (differently over time), as shown by time-varying $E \rightarrow H^3$ in the box. }
\end{itemize}

Causally speaking, $H^1$ has static causal relations to the click; $H^2$ has dynamic causal relations to the click; and $H^3$ is not causal feature interactions, but shows (spurious) correlations to the click, due to the confounding effect related to 
\zy{$E$ (e.g., the confounding effect brought by the backdoor path $ H^{3}\leftarrow E\rightarrow H^{2}\rightarrow Y$).}
As only static causal relations are invariant over time and exert stable effects on click generation, we treat $H^1$ as stable feature interactions.

Nevertheless, the ERM paradigm exploits all correlations to fit the click data. 
The correlations between feature interactions and clicks could come from three sources: static causal relations ($H^{1}\rightarrow Y$), dynamic causal relations ($H^{2}\rightarrow Y$), and confounding effects (\eg $H^{3} \leftarrow E \rightarrow H^{2} \rightarrow Y$).
As a result, ERM inevitably learns unstable feature interactions, which have dynamic causal relations or spurious correlations to the click. 
It will make the model generalize poorly for future serving, since the effects of unstable feature interactions would vanish and even reverse~\cite{IRM,liyunqi-JCDL,PDA,gao2022causal}.


As such, we revise the objective of the CTR task as capturing stable feature interactions. 
Towards this goal, we split the historical data chronologically into several periods, enabling the learning of feature interactions that are stable across these periods.
In the language of invariant learning~\cite{IRM}, these periods form a set of environments, which are heterogeneous due to temporal influences, \eg the influence of the COVID-19 lockdown.
Supposing the heterogeneity is large enough, only feature interactions with static causal relations to clicks are stably useful for CTR prediction across these environments. As such, a CTR model built in such a manner could learn stable feature interactions and generalize better to predict future clicks.   

\subsection{Disentangled Invariant Learning}

In this subsection, we first discuss the potential and challenge of applying invariant learning to the CTR task, and then present the \textit{Disentangled Invariant Learning} (DIL) method from three aspects: optimization objective, learning algorithm, and model architecture.


\subsubsection{Invariant Learning Solution} To achieve our goal, an ideal way is to conduct causal discovery, but it is hard in real-world scenarios due to the non-stationary, high-dimension, and partially unobservable challenges~\cite{kun-CDsurvey}. We propose to leverage invariant learning~\cite{IRM}, which could capture the environment-invariant correlations given multiple heterogeneous environments, to learn stable feature interactions. There are two considerations: 1) periods could directly represent different environments, and 2) among the three types of correlations discussed in Section~\ref{sec:rethink}, the correlations brought by static causal relations are environment-invariant, while the correlations brought by dynamic causal relations and spurious correlations vary across environments\footnote{
\zy{Considering the correlations brought by $H^2\rightarrow Y$ and confounding effects (\eg $H^3\leftarrow E \rightarrow H^2 \rightarrow Y$), 
as $E$ and the relations (\eg $H^2\rightarrow Y$) drift over time, }
these correlations vary over environments.}, \ie are environment-specific~\cite{HRM,liyunqi-JCDL}.
It is thus possible to learn stable feature interactions by capturing environment-invariant correlations with invariant learning.

However, existing invariant learning methods are not applicable to the CTR task, because the sufficiency condition of their basic assumption (Assumption~\ref{assump}) is not satisfied. 
\begin{itemize} [leftmargin=*]
    \item Clicks are affected by both feature interactions with static causal relations to clicks and feature interactions with dynamic causal relations to clicks, \ie clicks entangle environment-invariant and environment-specific correlations. That means clicks cannot be sufficiently predicted with only environment-invariant correlations, violating the sufficiency condition.
\end{itemize}
To overcome this challenge, the key lies in the disentanglement of environment-invariant and environment-specific correlations. With the disentanglement, the influence of environment-specific correlations could be eliminated from clicks, making the sufficiency condition satisfied. 
To this end, we propose a new disentangled invariant learning method for CTR models, which contains a new optimization objective, learning strategy, and model architecture.

\vspace{+3pt}
\subsubsection{Disentangled Invariant Learning Objective}\label{sec:dis_inv}
We first divide the model parameters for feature interaction modeling into environment-invariant part $\phi_{s}$ and environment-specific parts $\{\phi_{t}\}_{t}$ (short for ${\{\phi_{t}\}_{t=1}^{T}}$, and $\phi_{t}$ is used for $t$-th environment), to capture environment-invariant and environment-specific correlations, respectively. 
For example, we can equip \zy{the backbone model FM} with environment-invariant feature embedding $\bm{v}^s_{i}$ and environment-specific feature embedding $\bm{v}^t_{i}$ for any $i$-th feature, and then $\phi_{s} = \{\bm{v}^{s}_{i}\}_{i}$ and $\phi_{t} = \{\bm{v}^{t}_{i}\}_{i} $ (\cf \zy{ Section~\ref{sec:model_arch}}).
With both $\phi_s$ and $\phi_t$, we could sufficiently predict the click for a sample $(\bm{x},y) \in \mathcal{D}_{t}$ as follows:
\begin{equation}
\vspace{-1pt}\small \label{eq:emb}
      \hat{y} = f(\bm{x};\phi_{s},\phi_{t}),
\end{equation}
where $f(\bm{x};\phi_{s},\phi_{t})$ is the modified CTR model. We present the detailed model architecture in Section~\ref{sec:model_arch} later.


To learn  $\phi_{s}$ and $\{\phi_{t}\}_{t}$, on one hand, we need an invariant learning objective, to make $\phi_{s}$ capture the environment-invariant correlations. On the other hand, we need environment-specific objectives to force $\{\phi_{t}\}_{t}$ to capture the environment-specific correlations. To this end, we propose the following overall optimization objective:
\vspace{-1pt}
\begin{subequations}\label{eq:overall-obj} \small
\begin{align}
    min_{\phi_s} &
      \sum_{t=1}^{T} w_{t} R(\mathcal{D}_{t}; \phi_{s},\phi_{t}) + \lambda V_{R} \label{eq:over-a}\\
    \textit{s.t.} \quad & \{\phi_{s},\phi_{t}\} \in \underset{\{\phi_{s},\phi_{t}\}}{\mathrm{argmin}} \, R(\mathcal{D}_{t}; \phi_{s},\phi_{t}) + \eta L_{t}, \forall t, \label{eq:over-b}
\end{align}
\end{subequations}
where \eqref{eq:over-a} and \eqref{eq:over-b} correspond to the invariant learning objective and environment-specific objectives, respectively. \zy{We explain these learning objectives in detail below.} 

\vspace{3pt}
 \noindent  \textbf{- Invariant learning objective (Equation~\eqref{eq:over-a})}. After $\phi_{t}$ has captured the environment-specific correlations, fixing $\phi_{t}$ is equal to removing its influences on clicks. Then capturing the environment-invariant correlations can sufficiently predict the clicks without the influence of environment-specific correlations, 
 \ie the sufficiency condition can be satisfied. Thus we fix the environment-specific $\phi_{t}$ to modify original invariant learning objective in Equation~\eqref{eq:IL} for learning $\phi_{s}$ as Equation~\eqref{eq:over-a},
    in which $R(\mathcal{D}_{t}; \phi_{s},\phi_{t})$ represents the loss on the $\mathcal{D}_{t}$ collected from $t$-th environment and is computed with the \textit{logloss}~\cite{autofis} as follows:
    \vspace{-1pt}
    \begin{equation*}
    \setlength{\abovedisplayskip}{4pt}
    	\setlength{\belowdisplayskip}{4pt}
        \begin{split}
            R(\mathcal{D}_{t}; \phi_{s},\phi_{t}) = & \sum_{(\bm{x},y)\in\mathcal{D}_{t}}  - y\cdot log\left(\sigma(f(\bm{x};\phi_{s},\phi_{t}))\right) \\
            &- (1-y)\cdot log(1-\sigma(f(\bm{x};\phi_{s},\phi_{t}))),
        \end{split}
    \end{equation*}
    where $\sigma(\cdot)$ denotes the sigmoid function. Besides, in Equation~\eqref{eq:over-a}, $V_{R}$  refers to the variance of $R(\mathcal{D}_{t}; \phi_{s},\phi_{t})$ over environments, computed like Equation~\eqref{eq:var}; 
    and $w_{t}$ represents the weight for $t$-th environment, which is computed with the softmax function as follows:
    \begin{equation} \small
    \setlength{\abovedisplayskip}{6pt}
    	\setlength{\belowdisplayskip}{6pt}
       w_{t} = \frac{e^{ R(\mathcal{D}_{t}; \phi_{s},\phi_{t})}}{\sum_{t^{\prime}=1}^{T} e^{ R(\mathcal{D}_{t^{\prime}}; \phi_{s},\phi_{t^{\prime}})}},
    \end{equation}
    where the temperature parameter of the softmax function is used but omitted here. 
   As such, we assign a higher weight to the environment with higher loss, considering that 
   paying more attention to difficult environments is helpful for improving the cross-environment performances~\cite{focalloss,groupdro} to learn stable feature interaction.
    
    
\vspace{3pt}
\noindent\textbf{- Environment-specific objectives (Equation~\eqref{eq:over-b})}. Combining $\phi_{s}$ and $\phi_{t}$ should fit $\mathcal{D}_t$ well, since we need the clicks can be sufficiently predicted with $\phi_{s}$ and $\phi_{t}$.  
Thus, for each environment $t$, we propose the optimization objective in Equation~\eqref{eq:over-b}, \ie
    \begin{equation*}\small
        min_{\phi_{s},\phi_{t}} R(\mathcal{D}_{t}; \phi_{s},\phi_{t}) + \eta L_{t},
    \end{equation*}
    where  $\eta$ is a hyper-parameter to control the strength of $L_{t}$, and $L_{t}$ is a regularizer to prevent
    $\phi_{t}$ from capturing the environment-invariant correlations and is only used for learning $\phi_{t}$.
    Formally,
    \begin{equation}\small
    \label{eq:lint}
        L_{t} = \sum_{t^{\prime}=1, t^{\prime} \neq t}^{T} \left [R(\mathcal{D}_{t^{\prime}};\phi_{s},\phi_{t^{\prime}})-R(\mathcal{D}_{t^{\prime}};\phi_{s},\phi_{t}) \right ],
    \end{equation}
    where $R(\mathcal{D}_{t^{\prime}};\phi_{s},\phi_{t})$ represents the loss when predicting clicks in $\mathcal{D}_{t^{\prime}}$ forcibly using $\phi_{t}$.  
    The key consideration is to make $t$-th environment-specific $\phi_{t}$ not contribute to fitting data collected from other environments $\{t^{\prime}|1\le t^{\prime}\le T, t^{\prime}\in \mathbb{N}, t^{\prime} \neq t\}$. 

The two optimization objectives are not isolated. The first objective is based on the second, since it requires $\phi_{t}$ has captured environment-specific correlations. Thus we merge them to form the overall optimization objective in Equation~\eqref{eq:overall-obj}. We term the overall optimization objective --- disentangled invariant learning objective.

\vspace{-3pt}
\subsubsection{Learning Strategy} 
\label{sec:algDIL}

We now consider how to optimize the disentangled invariant learning objective (Equation~\eqref{eq:overall-obj}). 
\begin{algorithm}[t]
	\caption{DIL}
	\LinesNumbered
	\label{alg:DIL}
	\KwIn{Historical data chronologically split into $T$ environments, \ie $\mathcal{D}=\{\mathcal{D}_{1},\dots,\mathcal{D}_{t},\dots,\mathcal{D}_{T}\}$.}
    \zy{Initialize $\phi_{s}$ and $\{\phi_{t}\}_{t}$}\; 
     
    \While{Stop condition is not reached}{
    Randomly sample a batch of data from $\mathcal{D}$\;
    	Normally update model parameters except $\phi_{s}$ and $\{\phi_{t}\}_{t}$, and then keep them fixed\;
    	Randomly sample an environment $t\in \{1,\dots,T\}$ \;
    	// Update $\phi_{s}$,  fixing all $\phi_{t}$\;
    	Compute $\widetilde{\phi}_{s}$ with Equation~\eqref{eq:meta-training}\;
    	Update $\phi_{s}$ according to Equation~\eqref{eq:meta-test}\;
    	// Update $\phi_{t}$, fixing $\phi_{s}$\;
    	Update $\phi_{t}$ according to Equation~\eqref{eq:update_t}\; 
    }
\end{algorithm}
As we need to remove the influences of environment-specific correlations on clicks by fixing $\{\phi_{t}\}_{t}$ for the invariant learning objective, we propose a learning strategy, which updates the environment-invariant $\phi_{s}$ and environment-specific $\{\phi_{t}\}_{t}$ alternately as follows:

\vspace{3pt}
\noindent\textbf{- Update $\phi_{s}$}. Fixing $\{\phi_{t}\}_{t}$, we need to learn $\phi_{s}$ such that it is optimal for both the invariant learning objective and environment-specific objective as shown in Equation~\eqref{eq:overall-obj}. However, this is a hard bi-level optimization problem. 
    Instead of directly solving it, we solve it in a meta-learning manner similar to~\cite{meta-irm}. 
    Specially, we take the MAML~\cite{MAML} to update $\phi_{s}$, which has two main steps:
    \begin{itemize}[leftmargin=*]
        \item Step 1. Meta training, which focuses on the environment-specific objective. We sample an environment $t$ and compute the loss of environment-specific objective for it, then conduct an update to get a temporary $\widetilde{\phi}_{s}$ as follows:
    \begin{equation} 
    \vspace{-1pt}
    \label{eq:meta-training}  
    \setlength{\belowdisplayskip}{3pt}
        \widetilde{\phi}_{s} =  \phi_{s} - \nabla_{\phi_{s}} R(\mathcal{D}_{t}; \phi_{s},\phi_{t}).
    \vspace{-1pt}
    \end{equation}
    
    \item Step 2. Meta testing, which focuses on the invariant learning objective (Equation~\eqref{eq:over-a}). We take the $\widetilde{\phi}_{s}$ to compute the loss of the invariant learning objective, and further update the $\phi_{s}$ according to the loss. Formally,
    \begin{equation}\small \label{eq:meta-test}
        \phi_{s} \leftarrow \phi_{s} - \nabla_{\phi_{s}} \left (\sum_{t^{\prime}=1}^{T} w_{t^{\prime}} R(\mathcal{D}_{t^{\prime}}; \widetilde{\phi}_{s},\phi_{t^{\prime}}) + \lambda V_{R} \right ).
        \vspace{-1pt}
    \end{equation}
    \end{itemize}
    Note that $V_{R}$ is also computed  with $\widetilde{\phi}_{s}$. In this way, it is easier to find a $\phi_{s}$ that is generally good for the two optimization objectives, because of taking into account how an update for an environment-specific objective affects the invariant learning objective.

\vspace{3pt}
\noindent\textbf{- Update $\phi_{t}$}. 
After updating $\phi_s$, we directly optimize the environment-specific objective with $\phi_s$ fixed, to update $\phi_{t}$ (of the sampled environment) as follows:
    \begin{equation} \label{eq:update_t}
        \phi_{t} \leftarrow  \phi_{t} - \nabla_{\phi_{t}} \left( R(\mathcal{D}_{t}; \phi_{s},\phi_{t}) + \eta L_{t}\right).
    \end{equation}


These two updates are iterated alternately until convergence, and learning rates are used but omitted here. Algorithm~\ref{alg:DIL} summarizes the detailed learning algorithm of DIL. In each iteration, we first randomly sample a batch of data and update all parameters except $\phi_{s}$ and $\phi_{t}$ with normal training loss (\eg logloss) on the data (lines 3-4). Then we randomly sample an environment $t$ (line 5), and update $\phi_{s}$ (lines 7-8). Last, we update $\phi_{t}$ (lines 10).
Note that, in each iteration, we only utilize the batch of data.

\vspace{-7pt}
\begin{figure}[t]
  \centering
  \includegraphics[width=0.99\linewidth]{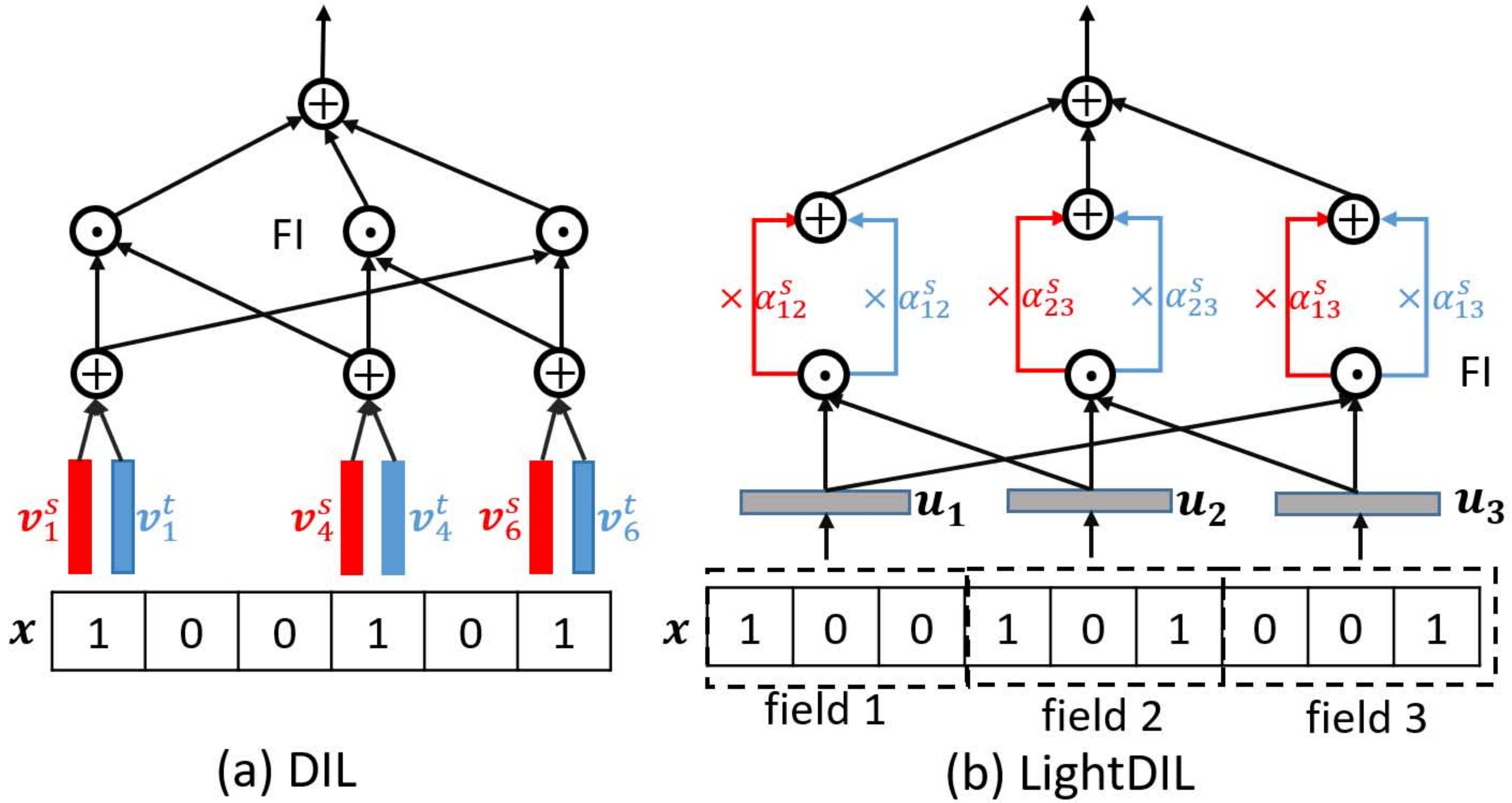}
  \vspace{-5pt}
  \caption{Model Architectures.  
(a) The embedding disentanglement model architecture for DIL. (b) The field-level disentanglement model architecture for LightDIL. "FI" denotes feature interactions. Red modules are environment-invariant, and the blue modules are environment-specific. }
\label{fig:models}
\vspace{-13pt}
\end{figure}


\vspace{+5pt}
\subsubsection{Model Architecture}
\label{sec:model_arch}
We now consider designing the model architecture of $f(\bm{x};\phi_{s},\phi_{t})$, taking FM as the base CTR model.

\vspace{3pt}
\noindent\textbf{- Embedding disentanglement}.  In light that the feature embedding is the core of feature interaction modeling \zy{ (especially for FM-based models)},
we first directly conduct feature embedding disentanglement. 
Specifically, for any $i$-th feature, we let it have an environment-invariant embedding $\bm{v}^{s}_{i}$ and a set of environment-specific embeddings $\{{\bm{v}^{t}_{i}}\}_{t}$, as shown in Figure~\ref{fig:models}(a). Then for a sample $(\bm{x},y)\in\mathcal{D}_{t}$ belonging to $t$-th environment, we modify FM (Equation~\eqref{eq:fm}) to generate the prediction as follows:
\begin{equation}\small
     \hat{y} = f(\bm{x};\phi_{s},\phi_{t}) = \sum_{i=1}^{N}\sum_{j>i}^{N} \langle \bm{v}^{s}_{i} + \bm{v}^{t}_{i}, \bm{v}^{s}_{j} + \bm{v}^{t}_{j} \rangle \cdot x_{i} x_{j},
\end{equation}
where $\phi_{s} = \{\bm{v}^{s}_{i}\}_{i}$ and $\phi_{t} = \{\bm{v}^{t}_{i}\}_{i} $.  This is the default disentanglement for the proposed Disentangled Invariant Learning (DIL)\footnote{\zy{There is another consideration for designing embedding disentanglement --- it is widely applicable since CTR models usually project features into embeddings.}}.


\vspace{3pt}
\noindent\textbf{- Field-level disentanglement}. 
The embedding disentanglement would result in $T$ times more model parameters
since each environment corresponds to a set of environment-specific embeddings, which dominates the model parameters in FM.  
What is worse, it is hard to train a model with massive parameters well.
This drives us to design a light version of DIL. 
Note that existing methods~\cite{autofis,FwFM} assign weights for feature interactions to model their importance, \ie the correlation strength to the click.
We could similarly assign environment-invariant and environment-specific weights for feature interactions, to capture the environment-invariant and environment-specific correlations, respectively. However, partial feature interactions could appear only in the future but not appear in history~\cite{FM}, resulting in their weights being unlearnable.  
To overcome the issue, we assign weights at the field level, as shown in Figure~\ref{fig:models}(b). Then we modify the FM in Equation~\eqref{eq:fm-field} as follows:
\begin{equation}\small \setlength{\belowdisplayskip}{3pt}
	\setlength{\abovedisplayskip}{3pt}
    \hat{y} = f(x;\phi_{s},\phi_{t}) = \sum_{i=1}^{M}\sum_{j>i}^{M} (\alpha^{s}_{ij} + \alpha^{t}_{ij}) \langle \mathbf{u}_{i}, \mathbf{u}_{j} \rangle,
\end{equation}
where $\alpha^{s}_{ij} \in \mathbb{R}$ ($\alpha^{t}_{ij} \in \mathbb{R}$) denotes the environment-invariant weight ($t$-th environment-specific weight) for the feature interaction between field $i$ and $j$, $\phi_{s} = \{ \alpha^{s}_{ij}\}_{ij}$, and $\phi_{t} = \{ \alpha^{t}_{ij}\}_{ij}$.
Note that feature embeddings do not belong to $\phi_{s}$ and ${\phi_{t}}$ here, and are normally updated (see Algorithm~\ref{alg:DIL}).
We name the DIL performing the disentanglement at the higher level of the feature field LightDIL. 

\vspace{3pt}
\noindent\textbf{- Inference.}  During inference, DIL and LightDIL only take $\phi_{s}$ to generate predictions, to pursue good generalization performance. 
For example, LightDIL generates the prediction as: $\hat{y} = f(x;\phi_{s},\varnothing) = \sum_{i=1}^{M}\sum_{j>i}^{M} \alpha^{s}_{ij} \langle \mathbf{u}_{i}, \mathbf{u}_{j} \rangle$. 
Another way is combining the $\phi_{s}$ learned in training 
and the $\phi_{t}$ estimated for a future period to generate predictions.
However, it is hard to forecast the $\phi_{t}$ of a future period. We leave it as future work.

\vspace{-0.2cm}
\section{EXPERIMENTS}
We conduct experiments to answer three research questions:

\begin{table*}[ht]
\renewcommand\arraystretch{0.93}
	\caption{Performance comparison between the baselines, DIL, and LightDIL. "Rel. Impr." denotes the relative improvement over FM \wrt AUC. The value on the left (right) of "$\pm$" represents the average (the standard deviation) \wrt the corresponding metric.  }
	\vspace{-10pt}
	\label{tab:overall}
	\centering
	\setlength{\tabcolsep}{1.2mm}{
		\resizebox{0.8\textwidth}{!}{
			\tiny
\begin{tabular}{c|ccc|ccc}
\hline
\multirow{2}{*}{Model} & \multicolumn{3}{c|}{Douban}                             & \multicolumn{3}{c}{ML-10M}                              \\
                       & AUC $\uparrow$      & logloss $\downarrow$ & Rel. Impr. & AUC $\uparrow$      & logloss $\downarrow$ & Rel. Impr. \\ \hline
FM                     & 0.7959 $\pm$ 0.0076 & 0.4459 $\pm$ 0.0269  & -          & 0.7158 $\pm$ 0.0081 & 0.4439 $\pm$ 0.0202  & -          \\
FwFMs                  & 0.8071 $\pm$ 0.0062 & 0.4262 $\pm$ 0.0217  & 1.41\%     & 0.7146 $\pm$ 0.0081 & 0.4465 $\pm$ 0.0202  & -0.17\%    \\
AFM                    & 0.8055 $\pm$ 0.0056 & 0.4271 $\pm$ 0.0207  & 1.21\%     & 0.7095 $\pm$ 0.0081 & 0.4579 $\pm$ 0.0167  & -0.88\%    \\
AutoFIS                & 0.8064 $\pm$ 0.0059 & 0.4247 $\pm$ 0.0201  & 1.32\%     & 0.7135 $\pm$ 0.0074 & 0.4488 $\pm$ 0.0208  & -0.32\%    \\
PROFIT                 & 0.8077 $\pm$ 0.0069 & 0.4248 $\pm$ 0.0216  & 1.48\%     & 0.7158 $\pm$ 0.0079 & 0.4498 $\pm$ 0.0216  & 0          \\
CFM                    & 0.8023 $\pm$ 0.0060 & 0.4289 $\pm$ 0.0208  & 0.80\%     & 0.7091 $\pm$ 0.0088 & 0.4473 $\pm$ 0.0151  & -0.93\%    \\
Group-DRO              & 0.8006 $\pm$ 0.0060 & 0.4290 $\pm$ 0.0205  & 0.59\%     & 0.7133 $\pm$ 0.0081 & 0.4443 $\pm$ 0.0186  & -0.35\%    \\
V-REx                  & 0.7996 $\pm$ 0.0065 & 0.4313 $\pm$ 0.0223  & 0.46\%     & 0.7112 $\pm$ 0.0081 & 0.4487 $\pm$ 0.0199  & -0.64\%    \\ \hline
DIL                    & 0.8081 $\pm$ 0.0058 & 0.4230 $\pm$ 0.0193  & 1.53\%     & 0.7162 $\pm$ 0.0082 & 0.4447 $\pm$ 0.0195  & 0.05\%     \\
LightDIL               & \textbf{0.8089} $\pm$ 0.0057 & \textbf{0.4219} $\pm$ 0.0193  & 1.63\%     & \textbf{0.7171} $\pm$ 0.0076 & \textbf{0.4438} $\pm$ 0.0198  & 0.18\%     \\ \hline
\end{tabular}
		}
  \vspace{-1pt}
\justify
\scriptsize{\zy{We have conducted a $t$-test, verifying the improvements of our best method to the best baseline are statistically significant on AUC (p-value $<0.05$).}}
	}
	\vspace{-0.3cm}
\end{table*}

\noindent\textbf{RQ1:} How does DIL perform on real-world data?  
\noindent\textbf{RQ2:} 
How do 
\zy{the design choices} of DIL
affect its performance?
\noindent\textbf{RQ3:} Can DIL learn stable feature interactions when there are spurious correlations? 
And can DIL learn stable feature interactions when there are dynamic causal relations (\ie the sufficiency condition of invariant learning is not satisfied)?

\vspace{-0.3cm}
\subsection{Experimental Settings}

\noindent\textbf{$\bullet$ Baselines.}
 We compare the proposed methods with the following recommender methods: 1) the basic FM~\cite{FM}; 2) FwFMs~\cite{FwFM}, which models the importance of feature interactions via field weights; 
\zy{3) AFM~\cite{AFM}, which utilizes an attention network to learn the importance of feature interactions;}
4) AutoFIS~\cite{autofis} and PROFIT~\cite{autodsn}, which are advanced neural architecture search (NAS) based methods for selecting more effective feature interactions; 5) CFM~\cite{liyunqi-JCDL}, which is a method for learning robust feature interactions (defined as causal feature interactions); 
In addition, we further compare DIL with two Out-of-Distribution (OOD) generalization methods, including one representative invariant learning method --- V-REx~\cite{krueger2021out} and a distributionally robust optimization method --- Group-DRO~\cite{groupdro}. \zy{For a fair comparison, we implement V-REx and Group-DRO based on FM.}
\zy{Considering that we aim at learning stably useful feature interactions to generalize well for the future, we select baselines 1) modeling the usefulness of feature interactions (FwFMs, AFM, AutoFIS, PROFIT, and CFM), or 2) pursuing the capability of OOD generalization (V-REx, Group-DRO, and CFM).}




\noindent\textbf{$\bullet$ Evaluation metrics and hyper-parameters.}
The widely used metrics for CTR prediction are Area under the ROC Curve (AUC) and logloss (\ie binary cross entropy). In this
work, to evaluate whether a model can generalize well to multiple future periods (testing environments), we report the average values of AUC and logloss over different testing environments. 

For a fair comparison, we learn all models based on the binary cross entropy loss and tune them according to the AUC metric computed on the validation environments.
Meanwhile, for all models, we set the feature embedding size as $48$ and take the Adam~\cite{adam} as the optimizer with the default batch size of $8,192$. We leverage the grid search to find the best hyper-parameters for each model.
\zy{In particular, we search the learning rate in the range of $\{1e\text{-}2, 1e\text{-}3, 1e\text{-}4\}$, and the $L_2$ regularization coefficient of the embedding layers in the range of $\{1e\text{-}1, 1e\text{-}2, \dots, 1e\text{-}7\}$ for all models. }
For the special hyper-parameters of baselines, we search most of them in the ranges provided by their papers.
\zy{For the special hyper-parameters  $\lambda$ and $\eta$ of our method (in Equation~\eqref{eq:overall-obj}), we search both of them in the range of $\{1e\text{-}1, 1e\text{-}2, 1e\text{-}3, 1e\text{-}4\}$.
}
\vspace{-2pt}
\subsection{Experiments on Real-world Data} 

 \subsubsection{Real-world Datasets}
 \zy{We conduct extensive experiments on the two representative datasets for CTR prediction: MoviveLens-10M and Douban, the statistics of which are summarized in Table~\ref{tab:statistics1}.
 }
 
\vspace{-2pt}
\noindent\textbf{- MovieLens-10M} (ML-10M)~\cite{movielens} is released by Grouplens Research, containing ratings \zy{collected from 1995 to 2009 with rich features belonging to various fields, such as age, gender, and movie category.} We only preserve the data collected from July 2002 to December 2008, considering the data sparsity of other years. \zy{ We then chronologically split the data into 13 periods by treating six months as a period. We further divide these periods into the training, validation, and testing environments according to the ratio of 5:4:4 with the order kept.}

\noindent\textbf{- Douban}\footnote{https://www.csuldw.com/2019/09/08/2019-09-08-moviedata-10m/.} is a popular dataset for CTR prediction, which contains the rating data \zy{collected from 2012 to 2019} on Douban Website. \zy{It contains user and item features belonging to the field of language, actors, \etc}~ \zy{Similarly, we preserve the data ranging from January 2012 to June 2019 and then split it into 15 periods with the first 5 for training, the middle 5 for validation, and the last 5 for testing.}

In both ML-10M and Douban, the ratings range from 1 to 5. Following~\cite{ wang2022causal}, we treat the user-item interactions with ratings $\ge3$ as positive samples with the click label $y=1$ (otherwise, $y=0$).  
Meanwhile, to ensure the dataset quality and avoid inadequate learning for extremely sparse features, we set the features appearing less than 10 times as a dummy feature (`other')~\cite{autofis,autodsn}.
 
 

 %

\begin{table}[t] 
\centering
\caption{Statistics of the evaluation datasets.}
\label{tab:statistics1}
\vspace{-6pt}
\setlength{\tabcolsep}{1.2mm}{
\resizebox{0.40\textwidth}{!}{
\begin{tabular}{ccccccc}
	\hline
	Dataset & \#Instances       & \#Fields & \#Features  & \#Train & \#Valid & \#Test  \\ \hline
	ML-10M  & 1.0$\times10^{7}$ & 7        & 145896        & 1.7$\times10^{6}$ & 1.9$\times10^{6}$ & 1.5$\times10^{6}$ \\
	Douban  & 4.2$\times10^{6}$ & 19       & 307919        & 7.9$\times10^{5}$ & 6.8$\times10^{5}$ & 1.4$\times10^{6}$ \\
	Avazu   & 4.0$\times10^{7}$ & 23       & 645185        & 2.0$\times10^{7}$ & 7.1$\times10^{6}$ & 1.3$\times10^{7}$ \\ \hline
\end{tabular}
}
}
\vspace{-0.5cm}
\end{table}

\vspace{-3pt}
\subsubsection{Overall Performance Comparison (RQ1)}
Table~\ref{tab:overall} 
summarizes the overall performance comparison between the baselines, DIL, and LightDIL 
on Douban and ML-10M datasets. From the table, we have the following observations.
\begin{table*}[ht]
	\centering
	\renewcommand\arraystretch{0.84}
	\caption{Results of the ablation studies over LightDIL on Douban and ML-10M.}
	\vspace{-0.23cm}
	\label{tab:ablation}
	
\setlength{\tabcolsep}{1.8mm}{
\resizebox{0.85\textwidth}{!}{
\tiny

\begin{tabular}{c|cc|cc}
\hline
\multirow{2}{*}{Model}               & \multicolumn{2}{c|}{Douban}                & \multicolumn{2}{c}{ML-10M}                 \\
                                     & AUC $\uparrow$      & logloss $\downarrow$ & AUC $\uparrow$      & logloss $\downarrow$ \\ \hline
w/o environment weights              & 0.8069 $\pm$ 0.0057 & 0.4233 $\pm$ 0.0200  & 0.7156 $\pm$ 0.0078 & 0.4451 $\pm$ 0.0198  \\
w/o variance-based regularizer       & 0.8075 $\pm$ 0.0060 & 0.4243 $\pm$ 0.0212  & 0.7140 $\pm$ 0.0074 & 0.4479 $\pm$ 0.0210  \\
w/o meta-learning                    & 0.8057 $\pm$ 0.0065 & 0.4250 $\pm$ 0.0213  & 0.7122 $\pm$ 0.0084 & 0.4454 $\pm$ 0.0148  \\
w/o disentanglement                  & 0.8068 $\pm$ 0.0058 & 0.4251 $\pm$ 0.0210  & 0.7107 $\pm$ 0.0085 & 0.4542 $\pm$ 0.0128  \\
w/o environment-specific regularizer & 0.8062 $\pm$ 0.0056 & 0.4255 $\pm$ 0.0203  & 0.7150 $\pm$ 0.0079 & 0.4439 $\pm$ 0.0190  \\
LightDIL (original)                   & 0.8089 $\pm$ 0.0057 & 0.4219 $\pm$ 0.0193  & 0.7171 $\pm$ 0.0076 & 0.4438 $\pm$ 0.0198  \\ \hline
\end{tabular}
}
}
\vspace{-0.3cm}
\end{table*}
\begin{itemize}[leftmargin=*]

   	\item LightDIL achieves the best performance on both ML-10M and Douban.
    DIL is slightly worse than LightDIL but still outperforms all baselines on most of the metrics. 
Compared with the FM, FwFMs, AFM, AutotoFIS, and PROFIT, which are learned with ERM, 
the better results of DIL indicate that learning stable feature interactions could bring better generalization for future serving than pursuing ERM on historical data. The results verify 1) the rationality of our reformulation for the CTR task and 2) the effectiveness of our methods in learning stable feature interactions.
Compared to V-REx and Group-DRO, better results of DIL verify that DIL can overcome the challenge --- the clicks cannot be sufficiently predicted with environment-invariant correlations --- to capture environment-invariant correlations, learning stable feature interactions in CTR prediction.

	\item FwFMs, AFM,  AutoFIS, and PROFIT, which learn more effective and discard less useful feature interactions, outperform FM on Douban, but fail to surpass FM and even cause worse performance on ML-10M\footnote{\zy{AFM and PROFIT beat FM on MovieLens in their papers, which are different from our results. This is because their training and testing data are identically distributed.}}.
	These models learn feature interactions with strong correlations to clicks in historical data.
	~They thus are possibly dominated by spurious correlations while ignoring the causal relations, leading to poor generalization performance. 
	ML-10M has fewer fields than Douban as shown in Table~\ref{tab:statistics1}
 , and thus the detrimental effect of mistakenly discarding causally useful feature interactions is more severe on ML-10M. 
	\item 
	V-REx and Group-DRO outperform FM on Douban, but not on ML-10M, verifying that directly applying OOD generation methods for recommendation is not effective. 
	For the invariant learning method V-REx, its basic assumption (Assumption~\ref{assump}) cannot be satisfied in the CTR task. 
	Group-DRO just focuses on the worst environment, sacrificing the overall performance~\cite{common-good}.

		\item 
		CFM, which aims to learn personalized causal feature interactions, shows inferior performance.
		We attribute its poor performance to inaccurate causal effects estimation and capturing dynamic causal relations.  On one hand, for de-confounding, CFM needs to learn balancing weights for all training samples, which is non-trivial due to the large size of our datasets (at least 50 times larger than the dataset used in the CFM paper). 
And it is hard to estimate the personalized causal effects well due to the sparsity of user data. 
On the other hand, CFM captures dynamic causal relations, which could lead to poor generalization performances. 

\end{itemize}

\begin{figure}
    \centering
    \includegraphics[height = 3.3cm]{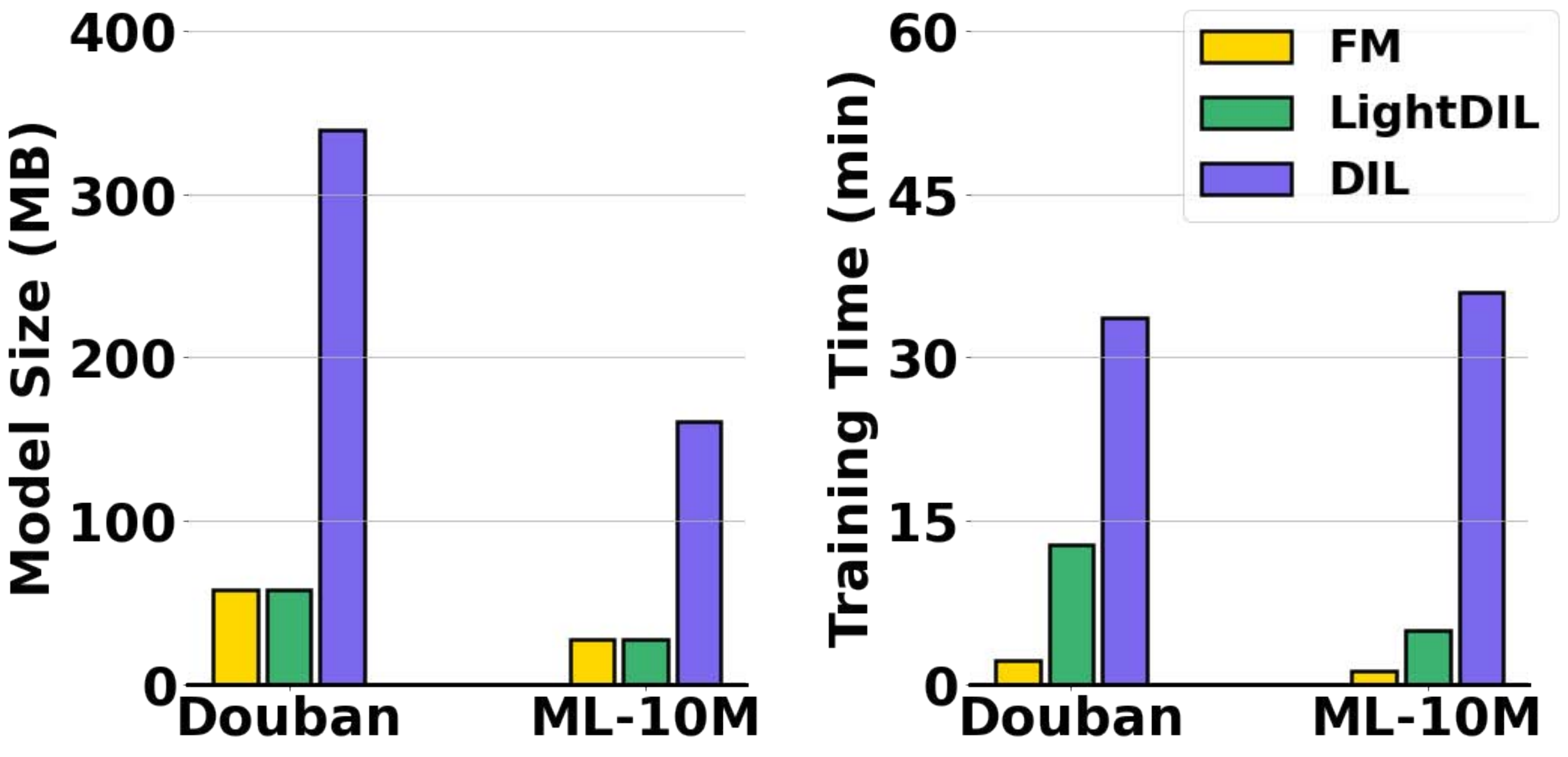}
    \vspace{-0.3cm}
    \caption{Comparison between FM, DIL, and LightDIL in terms of model size (left) and training time (right).} 
    \label{fig:cmpDIL}
    \vspace{-15pt}
\end{figure}

\vspace{-1pt}
\subsubsection{In-depth Analyses (RQ2)}
\label{sec:indepth}
\zy{
In this part, we first compare FM, DIL, and LightDIL on two datasets \wrt model size and training cost. We then conduct ablation studies on LightDIL to validate the effectiveness of different components of our method. Last, we study how the granularity of environment splitting affects the performance of our proposal based on LightDIL.
}


\vspace{2pt}
\noindent\textbf{$\bullet$ Modeling efficiency.}
We conduct a comparison between FM, DIL, and LightDIL regarding the model size and training time. As shown in Figure~\ref{fig:cmpDIL}, LightDIL has almost the same model size as FM, which is far smaller than that of DIL. Meanwhile, LightDIL could greatly reduce the time cost of training, compared to DIL. The results verify the validity of the field-level disentanglement in improving modeling efficiency\footnote{\zy{Although DIL's time (memory) cost seems not high (about 30 minutes or 300 MB) in our results, it is critical to improving the modeling efficiency since massive features/samples of industrial applications would immensely increase the cost. \textit{E.g.}, there could be billions of features~\cite{zhao2020distributed}, resulting in more than 1 TB parameters in DIL.}}.
Moreover, LightDIL improves the CTR performance over DIL, as illustrated in Table~\ref{tab:overall}. 
It reflects that the field-level disentanglement is sufficient to distinguish stable and unstable feature interactions with much fewer parameters. \zy{Regarding the performance difference between LightDIL and DIL, we attribute it to the discrepancy between them \wrt the disentanglement. For example, embedding disentanglement brings more model parameters for DIL while the corresponding feature interactions are more sparse (compared to the interaction between feature fields), making DIL hard to learn well. 
}
Regarding training, LightDIL is about five times slower than FM. 
This is because we adopt the paradigm of MAML~\cite{MAML} to update environment-invariant model parameters, which costs highly when computing the second-order gradients. In the future, it is possible to speed up LightDIL with the first-order approximation for MAML~\cite{MAML}.

\noindent\textbf{$\bullet$ Ablation studies.}
We next conduct ablation experiments to validate the effectiveness of the environment weights $w_t$ in Equation (\ref{eq:overall-obj}), the variance-based regularizer $V_{R}$ in Equation (\ref{eq:overall-obj}), the meta-learning module, the disentanglement, and the environment-specific regularizer $L_{t}$ in Equation (\ref{eq:lint}). 
Table~\ref{tab:ablation} enumerates the performance of the variants of LightDIL by disabling the above components one by one. From the table, we obtain the following findings.
\begin{itemize}[leftmargin=*]
    \item LightDIL w/o environment weights replaces the weights $w_t$ in Equation (\ref{eq:overall-obj}) with the uniform $\frac{1}{T}$.
    In Table~\ref{tab:ablation}, the performance gap between LightDIL and the variant w/o environment weights shows that using $w_t$ instead of equally 
treating each environment is helpful. The underlying reason is that using the weights will lead the model to focus more on the difficult environments during the optimization, validating the arguments in Section~\ref{sec:dis_inv}.
    
    \item The performance of LightDIL drops when we discard the variance-based regularizer, 
    reflecting its importance to balance the model's predictions over multiple environments. The variance-based regularizer can regulate the model to be robust across environments, helping to learn the environment-invariant correlations. 
    
    \item The variant w/o meta-learning is implemented by removing meta training in Equation (\ref{eq:meta-training}) and replacing $\widetilde{\phi_{s}} $ with $\phi_{s}$ in Equation (\ref{eq:meta-test}) to update $\phi_{s}$. 
    The inferior results of this variant indicate that meta-learning does help solve the challenging bi-level optimization problem in Equation (\ref{eq:overall-obj}) and enhance the CTR performance.

    \item Comparing LightDIL with the variant w/o disentanglement, we can observe that the disentanglement module effectively improves the CTR accuracy across environments. 
    The better result of LightDIL shows that the disentanglement is vitally important for capturing the environment-invariant correlations when the sufficiency condition (in Assumption~\ref{assump}) is not satisfied. 
    
    \item 
    We can find that the performance decreases when the environment-specific regularizer $L_{t}$ is disabled. The result verifies that $L_{t}$ could force $\phi_t$ to capture environment-specific correlations, and it is important for the disentanglement of $\phi_s$ and $\phi_t$.  
    
    \vspace{-4pt}
\end{itemize}

\begin{figure}[t]
    \centering
    \includegraphics[height=3.7cm]{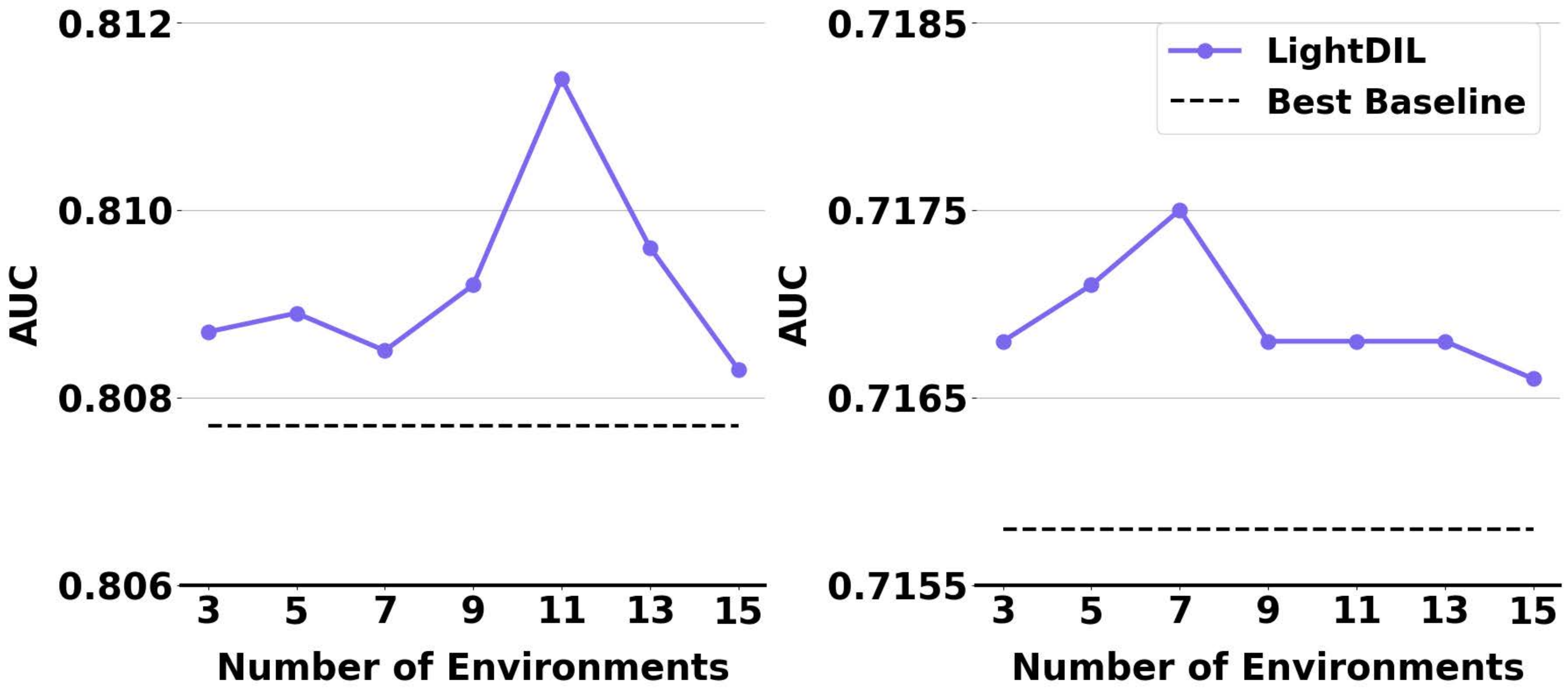}
    \vspace{-0.3cm}
    \caption{Performance of LightDIL on Douban (left) and ML-10M (right) when splitting training data into different numbers of environments. } 
    \vspace{-0.6cm}
    \label{fig:split-env}
\end{figure}
\vspace{2pt}
\noindent\textbf{$\bullet$ Granularity of separating environments.}
\zy{
We next study how the granularity of environment splitting affects the effectiveness of our proposal. We re-split the training data into $T$ environments and evaluate the performance of the LightDIL trained on the new split environments. We draw the performance curve of LightDIL when $T$ varies in $\{3,5,7,9,11,15\}$. Figure~\ref{fig:split-env} summarizes the results. We find that the performance of LightDIL increases first and then decreases as $T$ increases, \ie the granularity becomes small. We attribute it to that a more fine-grained splitting could increase the heterogeneity of training environments, benefiting the learning of the environment-invariant correlations. However, too fine-grained splitting would decrease the quality of training environments, \eg the data becomes very sparse, preventing the exhibition of environment-invariant correlations. Besides, we find LightDIL could beat the best baseline when $T$ varies in a wide range, verifying the superiority of our method. }




\subsection{Experiments on Semi-synthetic Data (RQ3)} %

\zy{We further design experiments on semi-synthetic datasets to verify whether DIL can successfully capture stable feature interactions by discarding the two types of environment-specific correlations: spurious correlations and correlations brought by dynamic causal relations, respectively. Note that if  dynamic causal relations exist, the sufficiency condition (Assumption~\ref{assump}) is not satisfied.}


\noindent\textbf{$\bullet$Semi-synthetic datasets.} 
It is hard to distinguish spurious correlations and dynamic causal relations in real-world datasets.
As such, we construct two semi-synthetic datasets based on \textbf{Avazu}, which is an advertising click dataset provided by Avazu corporation in the Kaggle CTR Prediction Contest\footnote{http://www.kaggle.com/c/avazu-ctr-prediction.}. The statistics of Avazu are provided in Table~\ref{tab:statistics1}. Avazu contains the click data between users and items spanning 10 days.
We aim to separately inject the spurious correlations and dynamic causal relations into two constructed datasets. And thus we hope the feature interactions in the original dataset are relatively stable.  
With this consideration, we select Avazu with the click data in a short period, where the correlations between the original feature interactions and the clicks are unlikely to shift. 
Thereafter, we treat each day as an environment and construct two datasets: Avazu with spurious correlations (SP-Avazu) and Avazu with dynamic causal relations (DC-Avazu), using the methods described in Appendix~\ref{sec:semi-syntheticdata}. 
For both SP-Avazu and DC-Avazu, we allocate the first 5 environments for training, the middle 2 for validation, and the last 3 for testing. 

\begin{figure}[t]
\centering
    \includegraphics[height=4.1cm]{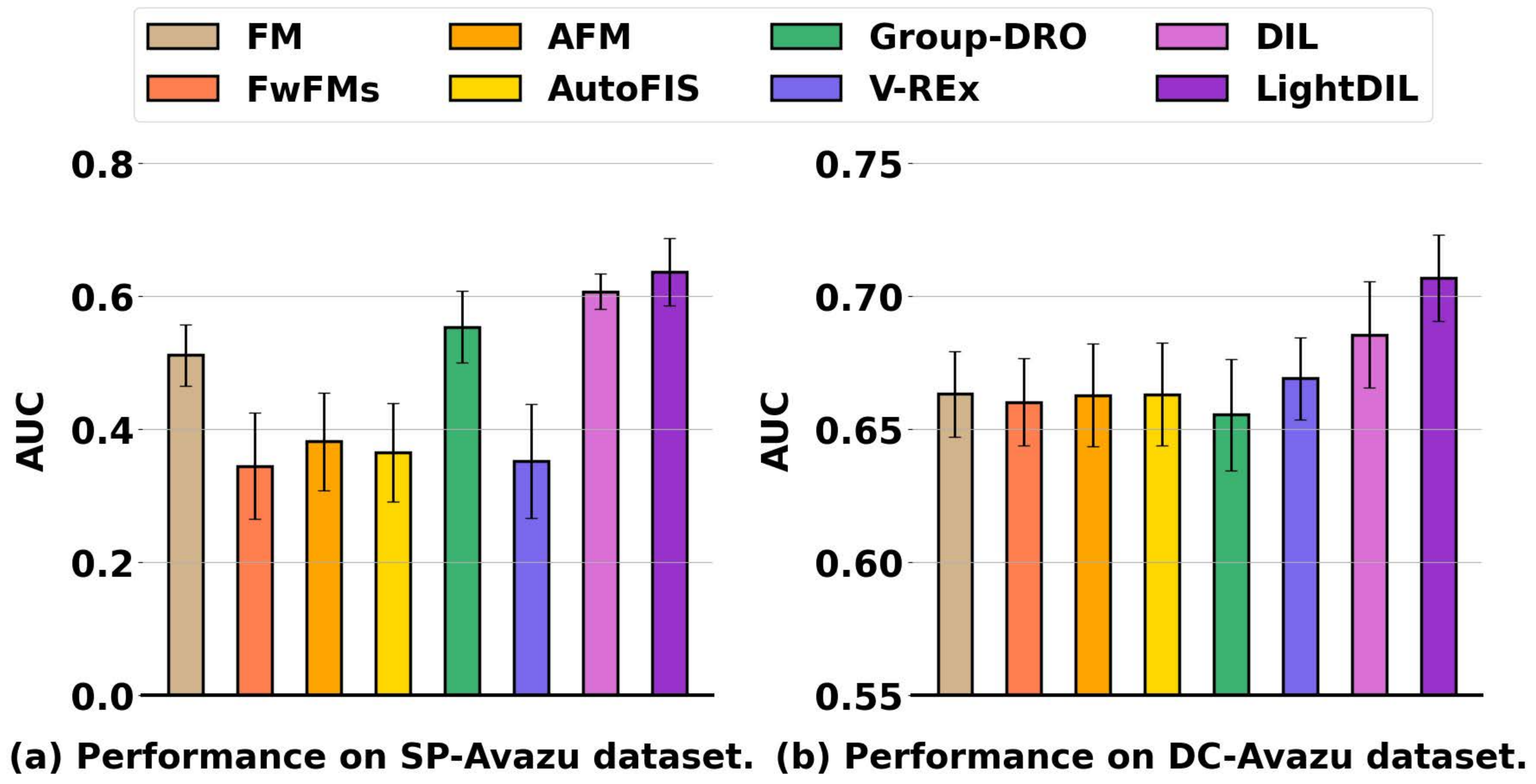}
    \vspace{-5pt}
    \caption{Performance of DIL, LightDIL, and the baselines on two semi-synthetic datasets: SP-Avazu and DC-Avazu.}
    \label{fig:SP-Avazu}
    \vspace{-15pt}
\end{figure}

\noindent\textbf{$\bullet$Performance comparison.} 
We conduct experiments on both SP-Avazu and DC-Avazu to verify that DIL could learn stable feature interactions with the existence of strong spurious correlations and dynamic causal relations. We compare our methods DIL and LightDIL with FM, FwFMs, AFM, AutoFIS, Group-DRO, and V-REx.
Here we ignore PROFIT and only compare with one representative NAS-based model AutoFIS because of the extremely long searching time of NAS-based models. 
Besides, we do not compare with CFM due to the lack of user ID information on Avazu, which is necessary for CFM. Figure~\ref{fig:SP-Avazu} illustrates the performance of these methods.
We can draw several conclusions from the figure.

\begin{itemize}[leftmargin=*]
    \item 
    On SP-Avazu, DIL and LightDIL substantially outperform the baselines. Moreover, they can generate meaningful predictions ($\text{AUC}>0.5$) while most baselines cannot achieve that.
    These results verify that the proposed disentangled invariant learning method can learn stable feature interactions and discard the feature interactions with spurious correlations. 
    
    \item On SP-Avazu, FM, FwFMs, AFM, and AutoFIS show poor performance with AUC close to or less than 0.5\footnote{\zy{
    That AUC is less than 0.5 is because the spurious correlations of training and testing environments are inverse.
    (\cf Appendix~\ref{sec:semi-syntheticdata}).
    }}. The results reflect that models built with the ERM paradigm are significantly affected by the feature interactions with strong spurious correlations.
    


    \item  On DC-Avazu, DIL and LightDIL also outperform all the baselines. Especially, DIL and LightDIL achieve better performances than V-REx on DC-Avazu. 
    This shows the effectiveness of the proposed method in learning stable feature interactions in the case where the sufficiency condition in Assumption~\ref{assump} is not satisfied.
    On SP-Avazu, we unexpectedly find that V-REx performs very poorly.
    The possible reason is that V-REx needs large enough distribution shifts between environments, which is not satisfied.
    The spurious correlations in training environments are mostly positive and thus the shifts might not be large enough.
    (see $\beta_{t}$ in Equation (\ref{eq:sp2}) in Appendix~\ref{sec:semi-syntheticdata}) 
    In contrast, the superior performance of DIL indicates that DIL can still learn stable feature interactions from the training environments with small shifts.

    
    

    
    \item 
    Regarding Group-DRO, it achieves good performance on SP-Avazu but performs poorly on DC-Avazu. Group-DRO focuses on the worst-case environment, which helps to avoid being overly reliant on spurious correlations. Nevertheless, ignoring some environments makes Group-DRO possibly capture the environment-specific correlations, bringing bad performance on DC-Avazu.

\end{itemize}
\vspace{-3pt}
To sum up, DIL and LightDIL can capture stable feature interactions and discard unstable feature interactions even when the sufficiency condition of invariant learning is not satisfied, enabling them to generalize well to unknown future periods.
\vspace{-5pt}

\section{Related Work}
%

\noindent\textbf{CTR prediction.}
CTR prediction has been widely studied for many years~\cite{ctrsurvey}. 
Since raw features rarely lead to satisfying results~\cite{ctrsurvey,min2022neighbour}, feature interaction modeling, which indicates the combination relationships of multiple features, becomes the focus of CTR prediction. 
FM is a pioneer and classical method, which models feature interactions in a factorization manner~\cite{FM}. 
Later, many methods are proposed to achieve more effective and complicated feature interaction modeling based on various neural networks, such as MLPs~\cite{DeepFM,can}, product-based neural networks~\cite{PNN,NFM}, attention-based networks~\cite{AFM,dualattention}, convolutional neural networks~\cite{FGCNN,CCPM}, and graph neural networks~\cite{gin,dualgnn}. Recently, NAS-based methods are proposed to automatically search the optimal network architecture for feature interaction modeling~\cite{nas-ctr,autoctr}, and automatically select/generate more effective feature interactions~\cite{autofis,autodsn,su2022detecting}. These methods reduce human efforts and achieve better feature interaction modeling. Besides, some methods~\cite{DIN,disenctr} additionally consider sequential user behavior modeling.

Although various CTR models are proposed, existing works are built in an ERM paradigm, blindly learning feature interactions useful for fitting historical clicks. 
On the contrary, we reformulate the CTR task to learn stable feature interactions and propose an invariant learning solution.
We notice that CFM~\cite{liyunqi-JCDL} is also not built with ERM. Differently, CFM aims to capture all causal feature interactions via distribution balancing, while we learn stable feature interactions (with static causal relations to click) instead of all causal feature interactions, using the invariant learning method.

\vspace{+2pt}
\noindent\textbf{Invariant learning.} 
To alleviate spurious correlations, previous studies have focused on invariant learning~\cite{IRM,HRM}. Generally speaking, invariant learning assumes that the training data is collected from distinct environments and aims to pursue robust predictions across multiple environments. 
In particular, Invariant Risk Minimization (IRM)~\cite{IRM} discovers stable features from multiple environments.
Following IRM, some studies pay attention to relaxing the linear assumption in IRM~\cite{ahuja2020invariant}, automatically splitting environments~\cite{creager2021environment, HRM}, and revising the regularizers~\cite{krueger2021out}. 
These methods assume the target could be sufficiently predicted with environment-invariant correlations, which is not satisfied in the CTR task. We extend invariant learning to
overcome this challenge.
Except for invariant learning solutions, Distributionally Robust Optimization improves the OOD generalization by minimizing the loss in the worst environment~\cite{dro-new,groupdro}. Stable learning instead achieves robust predictions via confounder balancing~\cite{shen2018causally}. 

Existing works have also considered invariant learning for recommendation debiasing ~\cite{inpf, he2020learning}, and alleviating spurious correlations in multimedia recommendation~\cite{du2022invariant}. SGL~\cite{he2020learning} focuses on dealing with selection bias. InvPref~\cite{inpf} infers the labels of heterogeneous environments and captures invariant preference across environments for debiasing in collaborative filtering. \zy{InvRL~\cite{du2022invariant} takes a similar method to  InvPref~\cite{inpf} but aims at alleviating spurious correlations in multimedia content}.  
\zy{Different from our proposal, 
these methods are not designed for the CTR task, do not learn stable feature interactions, and do not overcome the challenge that the sufficiency condition is not satisfied.}
\zy{
Besides, existing works have also considered achieving OOD generalization for recommendation by learning causal relations ( static and dynamic)~\cite{wang2022causal,causpref,liyunqi-JCDL}.
COR~\cite{wang2022causal} and CausalPref~\cite{causpref} both focus on learning the causal model of data generation. 
CFM~\cite{liyunqi-JCDL} is a CTR model and achieves OOD recommendations by distribution balancing as discussed before. Different from them, we aim at learning stable feature interactions with \textbf{static} causal relations to clicks. 
}

\vspace{-10pt}
\section{Conclusion}
In this work, we reformulate the CTR prediction for recommendation by learning stable feature interaction from split environments, aiming to generalize well for future serving. 
Towards this goal, we propose a novel disentangled invariant learning framework, which extends invariant learning to successfully capture invariant correlations when the prediction target entangles both environment-invariant and environment-specific correlations, \ie the sufficiency prediction condition is not satisfied. 
We conduct extensive experiments on both real-world and semi-synthetic datasets, providing insightful analysis for the effectiveness of our proposal.


This work showcased the limitation of the ERM learning paradigm for CTR prediction, despite its dominant role in recommendation research and industry.  We replace the ERM learning paradigm with the proposed disentangled invariant learning to learn stable information, achieving better generalization for future periods.  
We believe the proposed invariant learning paradigm could adapt to other real-world recommendation tasks and recommender models.  
In the future, we will apply our DIL to other state-of-the-art recommender models and other recommendation tasks (\eg collaborative filtering, sequential recommendation, and conversational recommendation~\cite{ren2022variational}).  Besides, we will also explore better disentanglement mechanisms for the proposed DIL.

\appendix
\section{Appendix}
\noindent\textbf{Construction of semi-synthetic datasets.}
\label{sec:semi-syntheticdata}
In this part, we present the details of how to generate the
two semi-synthetic datasets: SP-Avazu and DC-Avazu.

\noindent \textbf{$\bullet$ SP-Avazu}. \zy{To construct SP-Avazu with time-varying spurious correlations, we manually add a non-casual feature interaction correlated with the clicks.}
We inject the feature interaction by adding two identical binary features denoted as $x_{1}^{\prime}$ and $x_{2}^{\prime}$, considering that CTR models take raw features as inputs instead of feature interactions. 
Specifically, we obtain $x_{1}^{\prime}$ and $x_{2}^{\prime}$ by flipping the click label with environment-specific probabilities.
For a sample $(\bm{x},y)$ from $t$-th environment, we construct $x_{1}^{\prime}$ and $x_{2}^{\prime}$ as follows:
    \begin{equation}\small
    \vspace{-4pt}
    \begin{aligned}
    \epsilon_t\sim Bernoulli(p_t),\quad x_{1}^{\prime} = x_{2}^{\prime} = \epsilon_{t} (1-y)+(1-\epsilon_{t}) y,\\
    \end{aligned}
    \vspace{-3pt}
\end{equation}
where $p_t$ is the flipping probability in the $t$-th environment and we use $[0.1,0.15,0.2,0.25,0.5,0.5,0.7,0.8,0.85,0.9]$ for the ten environments of Avazu. 
Based on $p_t$, a flipping coefficient $\epsilon_t$ is drawn from the Bernoulli distribution, and then utilized to flip the click label. 
Thereafter, the synthetic feature interaction is the interaction between $x_{1}^{\prime}$ and $x_{2}^{\prime}$, and we obtain a constructed sample $(\bm{x}^{\prime},y)$, where $\bm{x}^{\prime} = [x_1,\dots,x_N,x_{1}^{\prime},x_{2}^{\prime}]$.

\noindent \textbf{$\bullet$ DC-Avazu}. It includes the dynamic causal relations from the synthetic feature interactions to clicks. Similar to SP-Avazu, we additionally inject two identical features $x_{1}^{\prime}$ and $x_{2}^{\prime}$, which causally revise the click label in environment-specific strengths.
Specifically, given a sample $(\bm{x},y)$ from the $t$-th environment, we first draw $x_{1}^{\prime}$ and $x_{2}^{\prime}$ from the Bernoulli distribution with probability $p_t$.
    Thereafter, we generate the new click label $\widetilde{y}$ by considering the new feature $x_{1}^{\prime}$ with the strength  $\beta_t$. 
    Formally, we have:
\begin{equation}\small
\label{eq:sp2}
\vspace{-4pt}
\begin{aligned}
    \epsilon\sim U(0,0.01), \quad x_{1}^{\prime}=x_{2}^{\prime}\sim Bernoulli(p_t),\\
    \alpha \sim U(0.55,0.65), \quad \widetilde{y}\gets\alpha y +\beta_t x_{1}^{\prime}x_{2}^{\prime}+\epsilon,\\
\end{aligned}
\vspace{-2pt}
\end{equation}
where $p_t$ for the Bernoulli distribution varies across environments. 
In this work, we use $[p_{1},\dots,p_{10}]=[0.1,0.2,0.2,0.2,0.2,0.1,0.2,0.2,\\0.2,0.2]$ for the ten environments of Avazu. The dynamic causal strength  $\beta_{t}$ varies in [0.6,0.5,0.15,0,-0.15,0,0.1,-0.15, -0.25, -0.4] for the ten environments. $\epsilon$ denotes the noises sampled from a uniform distribution to increase randomness. In a slight abuse of notation, since CTR prediction is a classification task, we treat samples with $\widetilde{y}>0.5$ as positive samples with rounded click labels $\widetilde{y}=1$ (otherwise, $\widetilde{y}=0$).
Finally, the synthetic feature interaction is the interaction between $x_{1}^{\prime}$ and $x_{2}^{\prime}$, and the newly constructed sample is $(\bm{x}^{\prime},\widetilde{y})$, where $\bm{x}^{\prime} = [x_1,\dots,x_N,x_{1}^{\prime},x_{2}^{\prime}]$. 

\begin{acks}
This work is supported by the National Key Research and Development Program of China (2021YFF0901603), the National Natural Science Foundation of China (62272437, 62121002), and the CCCD Key Lab of Ministry of Culture and Tourism. 
\end{acks}


\bibliographystyle{ACM-Reference-Format}
\balance
\bibliography{ref-bib}


\end{document}